\documentclass[a4paper,10pt]{article}
\usepackage[top=1in,bottom=1in,left=0.75in,right=0.75in]{geometry}
\usepackage [dvips]{graphicx}
\usepackage{amssymb,color,amsmath,fancyhdr,setspace,caption}

 \title{Interacting elephant random walks}
 \def\shorttitle{Interacting elephant random walks}
 \author{Chikashi Arita$^1$ and Eric Ragoucy$^2$} 
\def\shortauthor{C Arita \& E Ragoucy} 
 \def\address{
 1 Theoretische Physik, Universit\"at des Saarlandes, 66041 Saarbr\"ucken, Germany\\
 2 LAPTh, CNRS et USMB, 74940 Annecy, France}
 \def\abst{
The elephant random walk is a history-dependent random walk. We study a class of interacting elephant random walks.  Our model includes the exclusion process as a special case. By means of Monte Carlo simulations and mean-field arguments, we find that it exhibits a condensation phenomenon with a phase transition that is manifestly of first order. The transition point depends on the initial configuration. 
 }
 \date{\today}
\makeatletter
\def\@maketitle{ 
\begin{center} 
 \let \footnote \thanks
 {\Large\linespread{1.2}\selectfont\textbf{\@title}\par} \vskip 10mm 
 {\Large \@author} \vskip 5mm 
 {\address} \vskip 5mm 
 \textbf{Abstract} \end{center}
 \begin{quote} \abst \end{quote} \vskip 5mm 
\noindent\makebox[\linewidth]{\rule{\textwidth}{0.5pt}}}
\makeatother

\fancypagestyle{titlepage}{
 \lhead{} \chead{} \rhead{}
 \lfoot{} \cfoot{\thepage} \rfoot{} }

\begin{document}

\maketitle 

\thispagestyle{titlepage}

\section{Introduction}

A random walk is a fundamental concept to model the motion of a particle in a noisy environment. In one dimension and in discrete time, the position of a random walk is recursively given by $ X_t = X_ {t-1} + \sigma_t $ with $ \sigma_t = \pm 1 $. A value of $ \sigma_t $, which corresponds to a jump at time $t$, is stochastically given. For example, we assign  $ \sigma_t = +1 $ at probability $p$ and $ \sigma_t = -1 $ at probability $ q = 1- p $. The mean position satisfies the relation 
\begin{align}
 \langle X_t \rangle = \langle X_ {t-1} \rangle + \langle \sigma_t \rangle 
 = \langle X_ {t-1} \rangle + 2 p -1 ,
\end{align}
 which has the solution 
\begin{align}
 \langle X_t \rangle = X_0 + ( 2 p - 1 ) t . 
\end{align}
For $ p \neq 1/2 $, the walker exhibits a \textit{ballistic} behavior with velocity $ 2 p - 1 $. For $ p = 1/2 $, we consider the mean-square displacement $ M_t = \langle ( X_t - X_0 )^2 \rangle $, and we can show a \textit{diffusive} behavior $ M_t = t $.

The authors of \cite{bib:ST} introduced rightward and leftward hopping probabilities which depend on the history of the walker. The initial direction $ \sigma_1 $ is given randomly. The value of $ \sigma_t $ for $ t\ge 2 $ is determined as follows. We pick up one element $ \sigma_s $ from $ ( \sigma_1,\dots,\sigma_{t-1} ) $. If $ \sigma_s=+1 $, the walker jumps rightward i.e. $ \sigma_t = +1 $ (resp. leftward i.e. $ \sigma_t = - 1 $) with probability $r $ (resp. $ 1-r $). If $ \sigma_s=-1 $, the walker jumps rightward (resp. leftward) with probability $ 1-r $ (resp. $ r $). This model is called the elephant random walk. (See also a continuous-time version \cite{bib:PE}.)

The authors of \cite{bib:ST} derived a recursion relation for the mean position 
of the elephant 
\begin{align}
 \langle X_t \rangle = X_0 
 + \bigg( 1 + \frac{ 2r-1 }{ t-1 } \bigg) \Big( \langle X_{t-1} \rangle - X_0 \Big) .
\end{align}
Solving this relation, one finds the asymptotic behavior as 
\begin{align}
 \langle X_t \rangle = X_0 + \frac{ \sigma_1 \Gamma(t+2r-1)}{ \Gamma( 2r ) \Gamma(t) } 
 \simeq X_0 + \frac{ \sigma_1 }{ \Gamma( 2r ) } t^{ 2r-1 } \quad (t\to\infty) .
\end{align}
For $ r < 1/2 $, the mean displacement vanishes for large $t$ algebraically, 
in other words, the elephant tends to stay where it started.
In contrast, for $ r > 1/2 $, the elephant escapes from the starting point,
and the direction is determined by the first random decision.
In \cite{bib:ST}, the asymptotic behavior of the mean squared displacement for the elephant is also shown: 
\begin{align}
 M_t \simeq
 \frac{ t }{ 3-4r } \ ( r<3/4 ), \quad 
 t\ln t \ ( r=3/4 ), \quad 
 \frac{ t^{4r-2} }{ (4r-3) \Gamma (4r-2) } \ ( r>3/4 ) . 
\end{align}

Beyond the one-particle problem, it is important to consider many-particle systems with interactions in non-equilibrium statistical physics. The exclusion process on a one dimensional lattice is a basic example of stochastic interacting particle systems \cite{bib:MGP,bib:Spitzer}. Particles try to hop rightward and leftward with rates, say, $ p $ and $ q $, respectively. However, each site of the lattice can be occupied by at most one particle. Therefore an attempt of hopping is rejected, if another particle already occupies the target site. The exclusion process exhibits many interesting phenomena, such as shock and rarefaction waves, boundary-induced phase transitions (see e.g. \cite{bib:KRB-N}). It is used to model traffic flow, motion of molecular motors, surface growth, and many other systems \cite{bib:CMZ}. 

One of significant features of the exclusion process is its solvability (see e.g. \cite{bib:Schuetz}). It is known that the stationary state of the simple exclusion process is given by a product measure in the periodic boundary condition. In other words, the particles are uniformly distributed for a given density $ \rho $. Due to this fact, when the system size is large enough, the current and velocity are given by $ ( p - q ) \rho(1-\rho) $ and $ ( p - q ) ( 1-\rho ) $, respectively. 

In this paper we introduce a system of interacting elephant random walks, satisfying the exclusion principle: each particle has history-dependent transition rates which are similar to the original elephant random walk, but a jump is rejected when a target site is occupied. The system has three parameters $ p_\pm $, $ p_0 $ characterizing the directions of elephants' movements, and includes the symmetric and asymmetric simple exclusion processes, as we consider special sets of the parameters. When the number of particles is one (and $ p_+=1-p_- $), our model is equivalent to the original elephant random walk. We also remark that interacting elephants can choose \textit{staying} when one of the nearest-neighbor sites is already occupied by another particle, and they must stay when both two nearest-neighbor sites are already occupied. However, our model is of different nature from the single elephant random walk with stops introduced in e.g. \cite{bib:CSV, bib:KHL}. Since we impose the periodic boundary condition, the density $ \rho $ is conserved and considered as an additional parameter. By using Monte Carlo simulations, we explore global behaviors of the system.

 We shall find that there are two different regimes depending on the four parameters: a uniform phase and a condensed phase. For the former case, a mean-field analysis predicts the current of particles very well. On the other hand, for the latter case, a cluster is formed, and the standard mean-field theory fails.
 
 Instead of the full determination of the phase diagram in the four-dimensional space, we shall impose some restrictions on the four parameters and show plots of the current and other quantities vs one variable. We conjecture that the transition between the uniform and condensed phases is of the first order, i.e. the plots indicate discontinuity at some point. We shall also find that the transition point depends on the initial configuration of elephants. 

This article is organized as follows. In Section \ref{sec:model} we precisely define the model. In Section \ref{sec:mean-field} we perform a mean-field analysis to predict the particle current and other quantities. In Sections \ref{sec:11} and \ref{sec:01}, we show simulation results and check the validity of the mean-field formulas. We also discuss about the order of phase transitions and ergodicity breaking. In Section \ref{sec:discussions}, we give conclusions of this article and comments about existing studies on related models.

\section{Description of the model}\label{sec:model}
 \begin{figure} 
 \begin{center}
 \includegraphics[width=60mm]{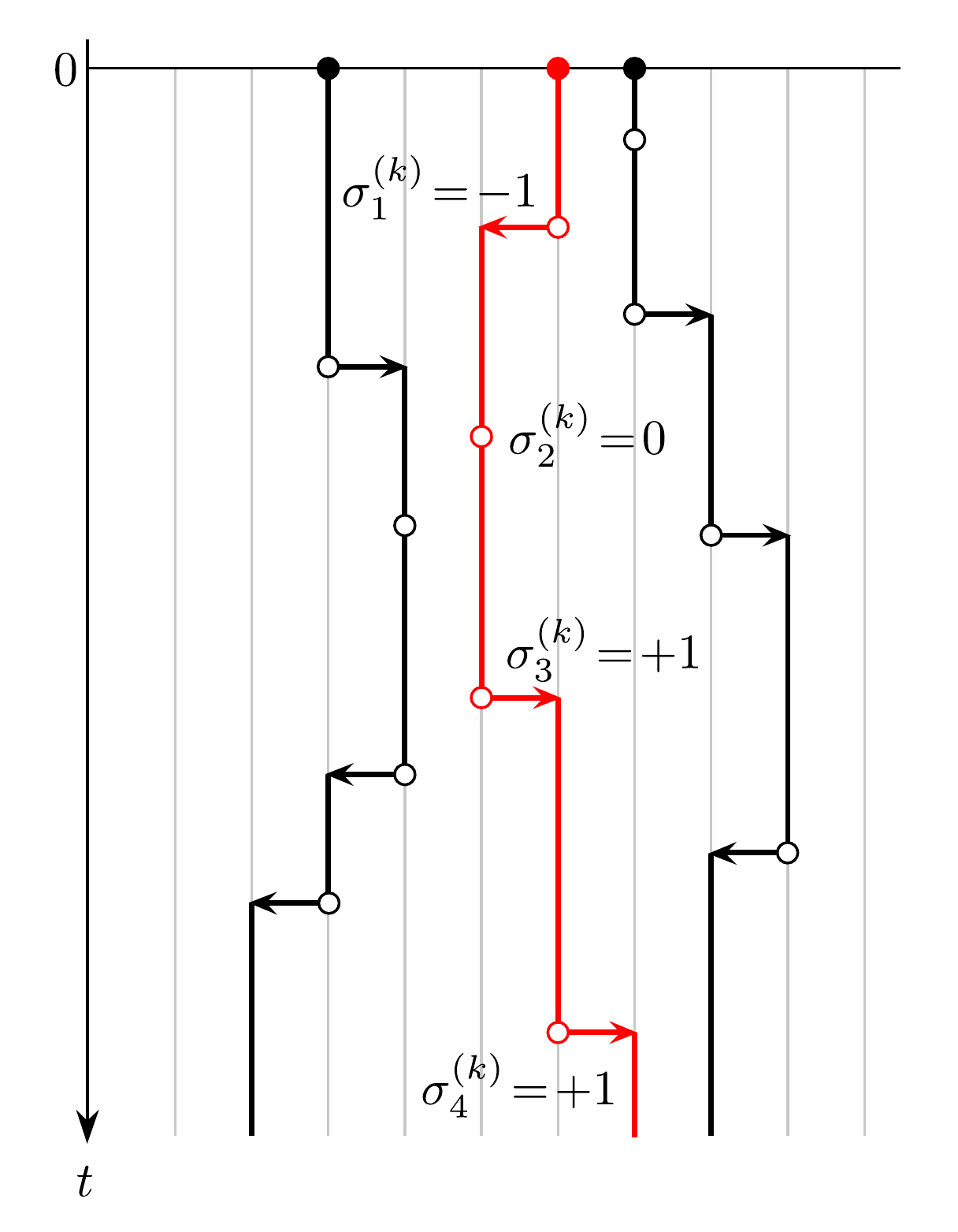}
 \end{center}
 \caption{\label{fig:illust}
 Illustration of trajectories of particles in our model. 
 Each site cannot be occupied by more than one particle, so each $ \circ$ corresponds an attempt to hop, which occurs at time $ t_n^{ (k) } $ so that $ \{ t_n^{ (k) } - t_{n-1}^{ (k) } \} $ obeys the exponential distribution with rate 1. The stochastic rule for the direction $ \sigma_n ^{ (k) } $ depends on the history, see the main text.}
\end{figure}

We consider an interacting particle system on a chain of $ L $ sites, where there are $N$ particles and periodic boundary conditions 
(i.e. the lattice is a ring). Each particle hops to a nearest-neighbor site, but each site is occupied by at most one particle, so a jump is allowed only when the target site is empty. These are basic rules in the exclusion process \cite{bib:MGP,bib:Spitzer}, see Fig.~\ref{fig:illust}. 
 In general, when one considers discrete-time dynamics in interacting particle systems, more than one particles can simultaneously attempt to move to the same site. In order to avoid this conflict, we rather use continuous-time dynamics. 
For each particle $k$, we generate a series of times $ \{ t_1^{ (k) }, t_2^{ (k) }, t_3^{ (k) }, \dots \} $, 
so that the difference sequence $ \{ t^{ (k) }_1-t^{ (k) }_0 , t^{ (k) }_2 - t^{ (k) }_1, t^{ (k) }_3 - t^{ (k) }_2, \dots \}$ $(t^{(k)}_0\equiv 0 )$ obeys the exponential distribution with rate 1, i.e.
the probability density function of $ t^{ (k) }_{n +1} - t^{ (k) }_n $ is given as 
 $f ( t^{ (k) }_{n +1} - t^{ (k) }_n = \tau ) = e^{ -\tau } $. 
 The time point $ t^{ (k) }_n $ ($n\ge 1 $) corresponds to the $ n $-th trial of hopping. 
The direction $ \sigma_n ^{ (k) } $ of the jump
 (rightward $+1$, leftward $-1$, or to `stay' $ 0 $)
 is stochastically determined, according to the following rule.

For the initial jump $n=1$, we choose one allowed direction $ \sigma^{ (k) }_1 $ from $ \{\pm 1 , 0\} $ randomly. 
 The direction of the second jump $ \sigma^{ (k) }_2 $ depends on $ \sigma^{ (k) }_1 $, the third one $ \sigma^{ (k) }_3 $ depends on $ ( \sigma^{ (k) }_1 , \sigma^{ (k) }_2 ) $, $\dots$, and for general $n$, $ \sigma_n^{(k)}$ depends on the history $ ( \sigma_1^{ (k) }, \sigma_2^{ (k) }, \dots ,\sigma_{n -1}^{(k)} ) $. We pick up $ m $ from $ \{1,2,\dots,n-1\} $ at the equal probability $ 1/(n-1) $, and we shortly write $ \sigma = \sigma_m^{(k)} $. When both neighbor sites of the particle $ k $ are empty, we choose $ \sigma_n^{(k)} = + 1 $ (resp. $ \sigma_n^{(k)} = - 1 $) with probability $ 0\le p_{ \sigma } \le 1$ (resp. $ q_{ \sigma } = 1- p_{ \sigma } $). If the left site is already occupied, the leftward jump is not allowed because of the exclusion principle, so $ \sigma_n^{(k)} = 0 $ with probability $ q_{ \sigma } $. Similarly, if the right site is already occupied, $ \sigma_n^{(k)} = 0 $ with probability $ p_{ \sigma } $. The particle has to stay $ ( \sigma_n^{(k)} = 0 ) $ with probability one when both neighbor sites are occupied. 
These hopping probabilities after choosing $ \sigma_m^{(k)} $are summarized as 
\newcommand{\occupied}{{\square\hspace{-2.25mm}\bullet\,}}
\newcommand{\boccupied}{{\square\hspace{-2.25mm}{\color{blue} \bullet}\,}}
\begin{align}
\label{eq:sigmakn-table}
\begin{array}{ccccc}
 \hline
 \sigma_n ^{ (k) } & \square\boccupied\square
 & \occupied\boccupied\square 
 & \square\boccupied\occupied
 & \occupied\boccupied\occupied \\ \hline\hline
 + 1 & p_\sigma & p_\sigma & 0 & 0 \\ \hline 
 -1 & q_\sigma & 0 & q_\sigma & 0 \\ \hline 
 0 & 0 & q_\sigma & p_\sigma & 1 \\ \hline
\end{array}
\end{align}
In the top row, the particle $k$ is in the middle site and colored in blue. It attempts to hop to one of the neighbor sites, but 
a hopping is allowed only when the target site is empty.

We have a convenient expression for $ p_\sigma $, in the case where $ \sigma$ only takes three values $ \{\pm 1, 0\} $: 
\begin{align}
 p_\sigma = 
 p_{+1} \frac{ \sigma ( \sigma + 1 ) }{2} 
 + p_{-1} \frac{ \sigma ( \sigma - 1 ) }{2} 
 + p_0 \big( 1 - \sigma^2 \big) 
 = p_0 + \frac{ p_{+1} - p_{-1} }{2 } \sigma 
 + \frac{ p_{+1} + p_{-1} - 2p_{0} }{2} \sigma^2 . 
\end{align}
Thanks to this formula, the probabilities of choosing
$ \sigma_n^{(k)} = \pm 1, 0 $ are eventually given by 
\begin{align}
\begin{array}{ccccc}
 \hline
 \sigma_n ^{ (k) } & \square\boccupied\square
 & \occupied\boccupied\square 
 & \square\boccupied\occupied
 & \occupied\boccupied\occupied \\ \hline\hline
 + 1 &P & P & 0 & 0 \\ \hline 
 -1 & Q & 0 & Q & 0 \\ \hline 
 0 & 0 & Q & P & 1 \\ \hline
\end{array}
\end{align}
where 
\begin{align}
\label{eq:P}
& P ( \sigma^{(k)}_1,\dots,\sigma^{(k)}_{n-1} ) 
= p_{0}
+ \frac{ p_{+1} - p_{-1} }{2(n-1)} \sum_{ m=1 }^{ n-1 } \sigma_m^{(k)}
+ \frac{ p_{+1} + p_{-1} - 2p_{0} }{2(n-1)} \sum_{ m=1 }^{ n-1 } \big[ \sigma_m^{(k)} \big]^2 , \\ 
& Q ( \sigma^{(k)}_1,\dots,\sigma^{(k)}_{n-1} ) 
 = 1- P ( \sigma^{(k)}_1,\dots,\sigma^{(k)}_{n-1} ) 
 = P ( \sigma^{(k)}_1,\dots,\sigma^{(k)}_{n-1} ) |_{p_\sigma \to q_\sigma (\sigma = \pm1,0)}. 
\end{align}
For example, when both neighbor sites of the particle $ k $ are empty,
$ \sigma_n^{(k)} = +1 $ is realized with probability $ P( \sigma^{(k)}_1,\dots,\sigma^{(k)}_{n-1} ) $,
and $ \sigma_n^{(k)} = -1 $ with probability 
 $ Q ( \sigma^{(k)}_1,\dots,\sigma^{(k)}_{n-1} ) $. 
The dynamics of elephants are essentially determined by the two quantities 
\begin{align}
 S^{ (k) }_n = \sum_{ m=1 }^{ n } \sigma_m^{(k)} ,
 \quad C^{ (k) }_n = \sum_{ m=1 }^{ n } \big[ \sigma_m^{(k)} \big]^2, 
\end{align}
as well as the configuration of the system. The full history is actually not needed. 

One of the most important quantities is the stationary current $ J $ in the study of interacting particle systems. 
Here we mention that the flip of the lattice orientation is a symmetry of our model: 
\begin{align}\label{eq:flip}
 J ( p_+, p_- , p_0 , \rho ) =& - J ( 1 - p_-, 1- p_+ , 1 - p_0 , \rho ) \\
 P ( \sigma^{(k)}_1,\dots,\sigma^{(k)}_{n-1} ) =& 1-
P ( -\sigma^{(k)}_1,\dots,-\sigma^{(k)}_{n-1} )\Big|_{\overset{\scriptstyle p_{\pm}\to 1-p_{\mp}}{ p_0\to 1-p_0}}
\end{align}
Note that the particle-hole duality 
is not a symmetry of the model in general: $ J ( p_+, p_- , p_0 , \rho ) \neq -J ( 1-p_+, 1-p_- , 1-p_0 , 1- \rho ) $. This is due to the fact that the jump probability of each particle 
is history-dependent but holes do not have a memory. 

\begin{figure}
 \begin{center}

\begin{align*}
\begin{matrix} 
 \textbf{(a)} & \textbf{(b)} \\ 
 \includegraphics[width=75mm]{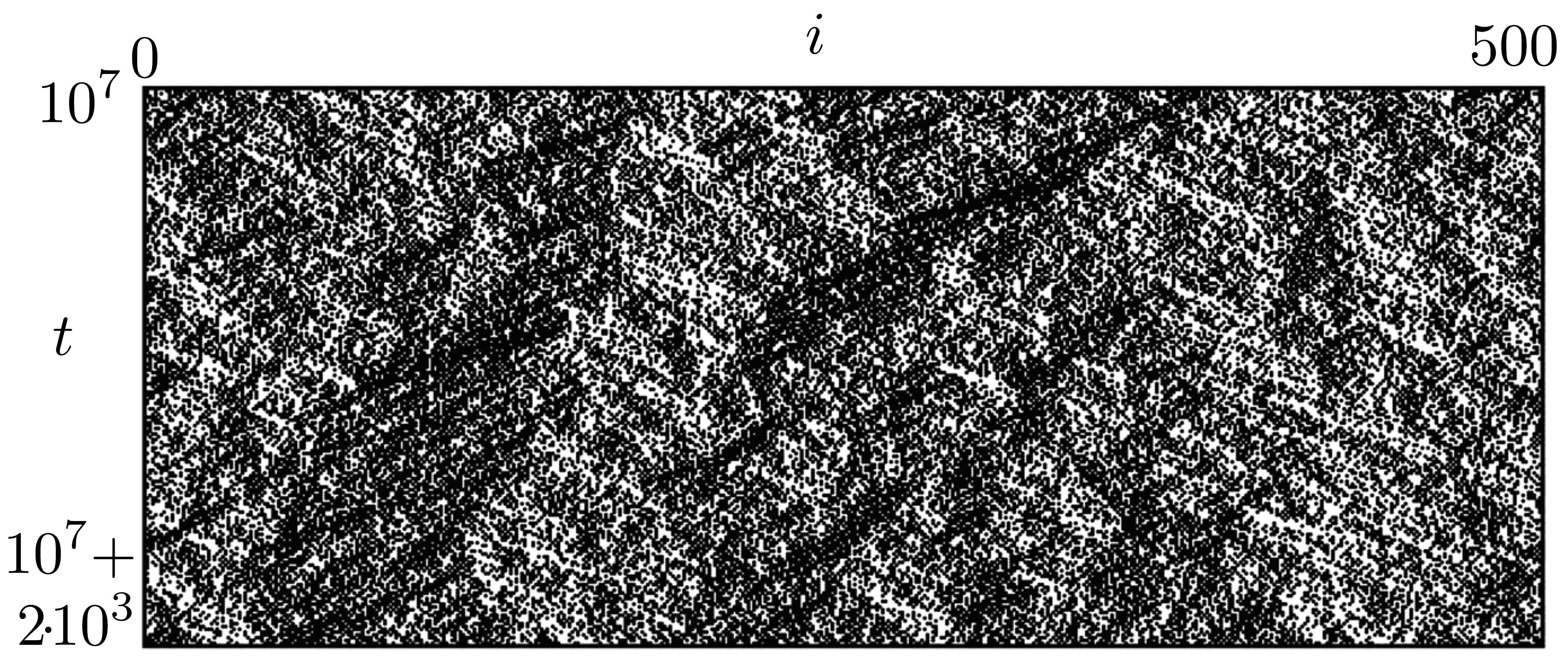} & \includegraphics[width=75mm]{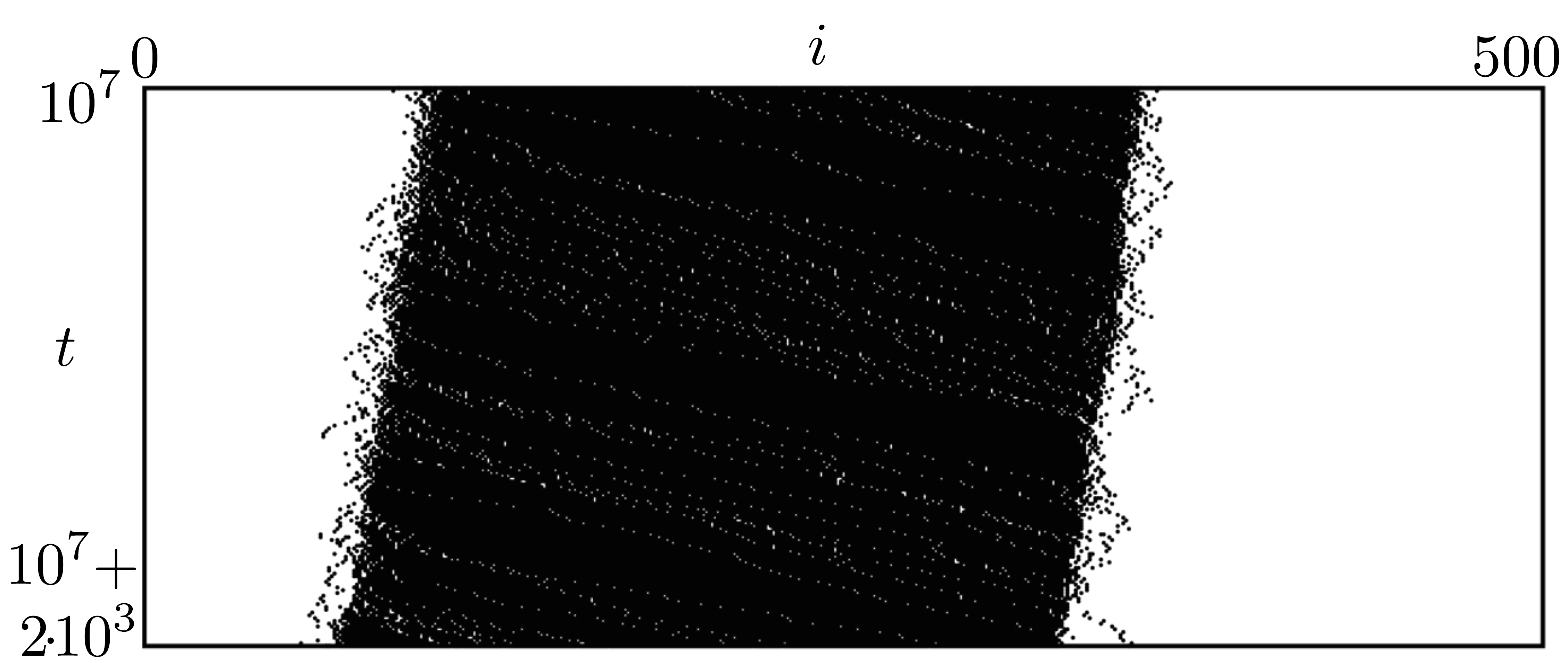} \\ 
 \textbf{(c)} & \textbf{(d)} \\ 
 \includegraphics[width=75mm]{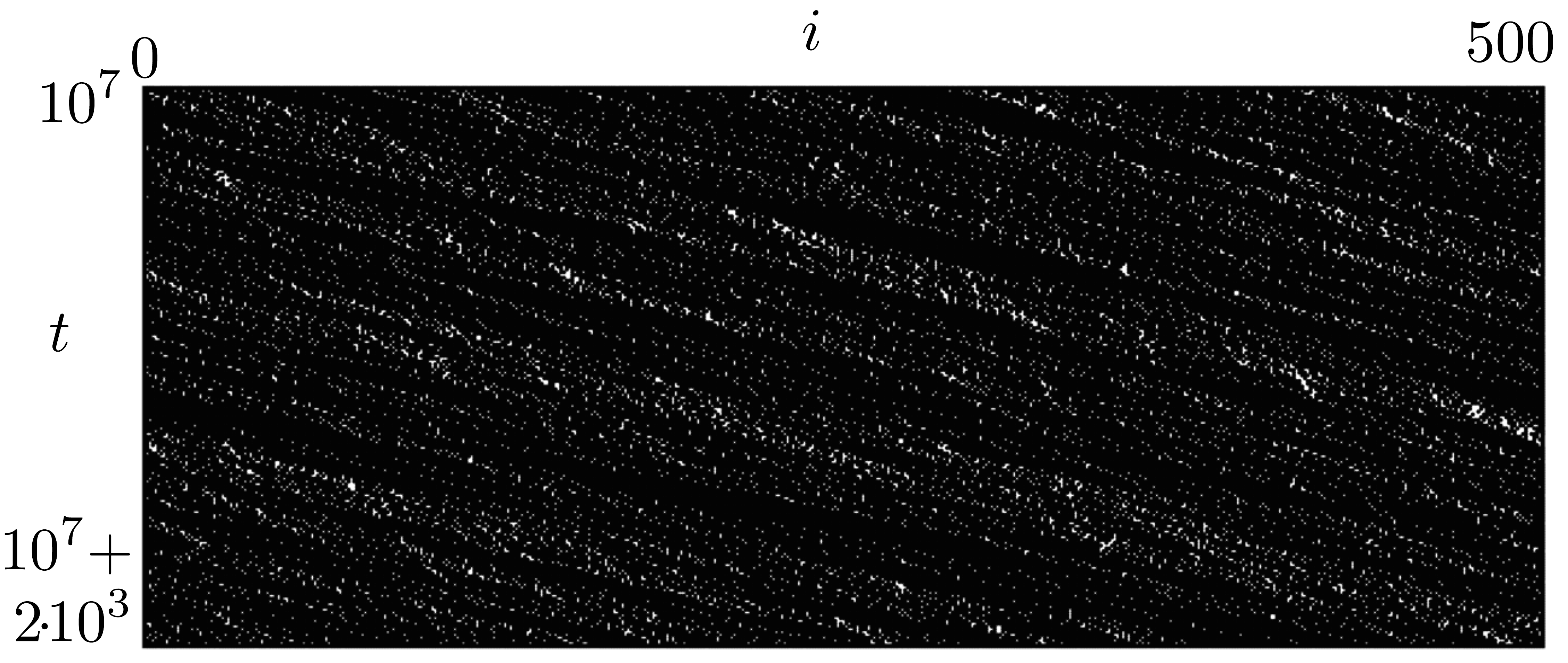} & \includegraphics[width=75mm]{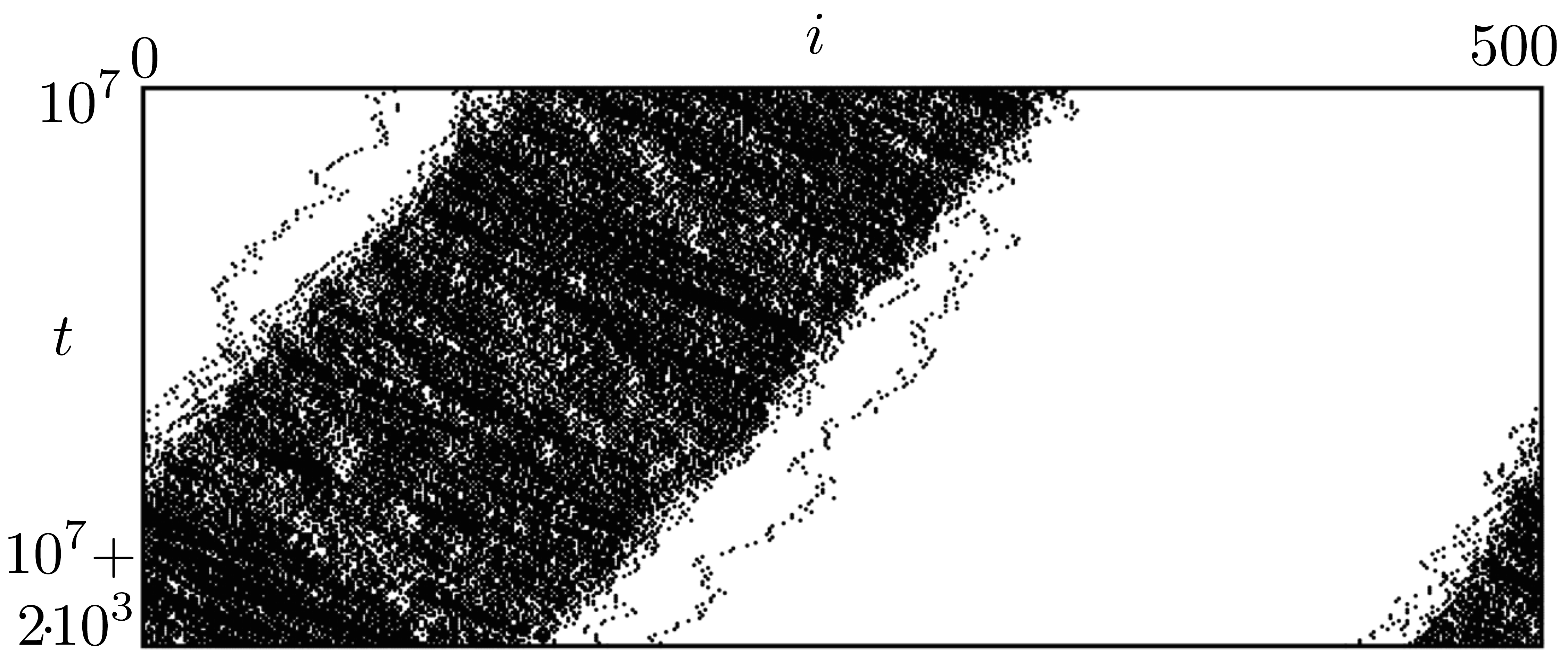}
\end{matrix} 
\end{align*}
 \caption{\label{fig:kymo}
 Kymographs for $ (p_+,p_-,p_0,\rho) = ( 1,1,0.75,0.5 ) $ (a), and $ ( 1,1,0.25,0.5 ) $ (b), 
 $ (p_+,p_-,p_0,\rho) = ( 0,1,0.2,0.9 ) $ (c), and $ ( 0,1,0.2 ,0.3) $ (d).
 The system size is $ L=1000 $.} 
 \end{center}
 \end{figure}

Let us look at kymographs, Fig~\ref{fig:kymo}, to get some insight on the model. Interestingly there are two completely different regimes. For the cases (a) and (c) we observe that the particles are distributed entirely over the chain (\textit{uniform phase}). On the other hand, for (b) and (d), a cluster is spontaneously formed (\textit{condensed phase}). In the next section we shall theoretically derive formulas for the current and other quantities by using mean-field assumption. We shall compare the theoretical predictions with simulation results, and discuss phase transitions between the two phases.

\section{Mean field analysis}\label{sec:mean-field}

\subsection{Global mean-field formulas}

Our model includes the usual simple exclusion process as a special case $ p_+ = p_- = p_0 $. For this case, the stationary state can be written as a product (Bernoulli) measure in the limit $ L\to \infty $.
 Let us assume the same distribution for the generic case of our model, in order to predict the stationary current $ J $.
For example the nearest-neighbor correlation takes a simple form
\begin{align}\label{eq:tt-mf}
 \langle \tau_{i}\tau_{i+1} \rangle = \rho^2 . 
\end{align}

 The quantity $ \frac{1}{n} S^{ (k) }_n $ is the effective probability 
of rightward hopping minus that of leftward hopping. 
Thus, within the mean-field assumption, we have 
\begin{align}
 \frac{1}{n} S^{ (k) }_n 
 = P (1-\rho) - Q (1-\rho) 
 = (2P-1) (1-\rho) . 
\end{align}
The quantity $ \frac{1}{n} S^{ (k) }_n$ is indeed regarded as the velocity of the elephant $ k $. 
On the other hand, the quantity $ \frac{1}{n} C^{ (k) }_n $ represents the probability that 
a jump is allowed. Therefore we have
\begin{align}
\label{eq:C-mf}
 \frac{1}{n} C^{ (k) }_n 
 = (1-\rho)^2 + P \rho(1-\rho) + Q \rho(1-\rho) = 1 - \rho . 
\end{align}
Substituting them into Eq.~\eqref{eq:P}, we find 
\begin{align}
 P = p_0 + \frac{ p_+ - p_- }{2} (2P-1) (1-\rho) + \frac{ p_+ + p_- - 2p_0 }{2}(1 - \rho ). 
\end{align}
Solving this self-consistency equation we get 
\begin{align}
\label{eq:P-mf}
 P = \frac{ p_- (1-\rho) + p_0 \rho }{ 1 - (p_+ - p_-) ( 1 - \rho ) } 
\end{align}
and 
\begin{align}
\label{eq:S-mf}
 \frac{1}{n} S^{ (k) }_n 
 = \frac{ ( p_+ + p_- ) ( 1 - \rho ) + 2 p_0 \rho - 1 }{ 1 - ( p_+ - p_-) (1-\rho) } (1-\rho ). 
\end{align}
The current is then predicted as 
\begin{align}
 J= ( P - Q ) \rho (1-\rho ) = (2P-1) \rho (1-\rho ) 
 = \frac{ ( p_+ + p_- ) ( 1 - \rho ) + 2 p_0 \rho - 1 }{ 1 - ( p_+ - p_-) (1-\rho) } \rho (1-\rho ). 
\label{eq:J-mf}
\end{align}
Note that $ J $ is measured at an arbitrary \textit{bond} between sites $i$ and $i+1$,
but it does not depend on $ i $ in a stationary state. 
We also remark that indeed we have a relation $ J= \rho \frac{1}{n} S^{ (k) }_n $.

The calculations above are of course not rigorous, so we have to check the validity of these formulas by comparing with simulations. 

\subsection{Local mean-field formulas}

 \begin{figure} 
 \begin{center}
 \includegraphics[width=100mm]{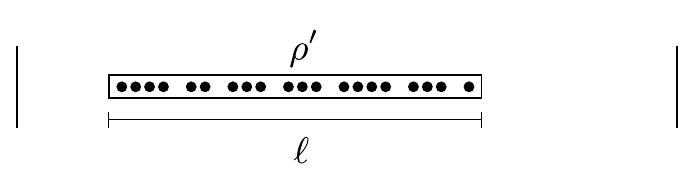}
 \end{center}
 \caption{
 Illustration for the local mean-field theory in the condensed phase. 
 There is a cluster of length $ \ell $. We denote the particle concentration by $ \rho' $ inside the cluster. 
 \label{fig:cluster}}
\end{figure}

We also introduce a mean-field theory for the condensed phase. We assume that, inside the cluster, the particles are uniformly distributed. Let us denote the length and the density of the cluster by $\ell $ and $\rho'=N/\ell$, respectively, see Fig.~\ref{fig:cluster}. Under this \textit{local} mean-field theory 
 we simply replace $ \rho $ by $ \rho' $
in the \textit{global} mean-field formulas \eqref{eq:C-mf} and \eqref{eq:S-mf} 
for $ \frac{C^{ (k) }_n}{n} $ and $ \frac{ S^{ (k) }_n}{n}$ 
because these are quantities for an individual particle: 
\begin{align}
 \frac{C^{ (k) }_n}{n} &= 1 - \rho' , \label{eq:C-lf}\\ 
 \frac{ S^{ (k) }_n}{n} &= \frac{ ( p_+ + p_- ) ( 1 - \rho' ) + 2 p_0 \rho' - 1 }{ 1 - ( p_+ - p_-) (1-\rho') } (1-\rho' ) , \label{eq:S-lf}
\end{align}
On the other hand, for the current $ J$ 
and the correlation function $ \langle \tau_{i}\tau_{i+1} \rangle $ 
we need to consider the two situations, 
i.e. the cluster covers the sites $ i $ and $ i+1 $ or not. 
Then we find 
\begin{align}
\nonumber
 J &= \frac{ \ell }{ L } 
 \frac{ ( p_+ + p_- ) ( 1 - \rho' ) + 2 p_0 \rho' - 1 }{ 1 - ( p_+ - p_-) (1-\rho') } \rho' (1-\rho' )
 + \frac{ L-\ell }{ L } 0 \\ 
& = \frac{ ( p_+ + p_- ) ( 1 - \rho' ) + 2 p_0 \rho' - 1 }{ 1 - ( p_+ - p_-) (1-\rho') } \rho (1-\rho' ) , \label{eq:J-lf}\\ 
 \langle \tau_{i}\tau_{i+1} \rangle &=
 \frac{ \ell }{ L } ( \rho' )^2 + \frac{ L-\ell }{ L } 0 = \rho \rho' . \label{eq:tt-lf}
\end{align}
These formulas contain the density $ \rho' $, but the determination of $ \rho' $ is nontrivial. 
It may depend on the parameters $ p_\pm, p_0 $ and $ \rho $ in general.

\clearpage

\section{Case $ p_+=p_-= 1 $. }\label{sec:11}

Now we show more quantitative simulation results. Our model contains four parameters; $ p_\pm $, $p_0 $ and $\rho $. It is pedagogic to impose restrictions on parameters, so that we can plot quantities such as $ J $ versus one variable. 
 
Let us begin with the case where $ p_\pm = 1 $. 
The probability \eqref{eq:P} is simplified as 
\begin{align} 
\label{eq:P:11p0}
 P = p_{0} + \frac{ 1- p_{0} }{ n-1 } C^{ (k) }_{n-1} . 
\end{align}
The quantity $ S^{ (k) }_{n-1} $ does not contribute to the hopping probabilities. 

When we further set $ p_0 = 1 $, our process is equivalent to the standard totally asymmetric simple exclusion process. For $ p_0 \neq 1 $, stationary states are non-trivial, and the uniqueness is no longer guarantied. Let us see whether the global mean-field theory gives a reasonable result.

\subsection{Uniform and condensed phases}

\begin{figure}[h]
 \begin{center} 
\begin{align*}
\begin{matrix} 
 \textbf{(a)} & \textbf{(b)} & \textbf{(c)} & \textbf{(d)} \\ 
 \includegraphics[width=40mm]{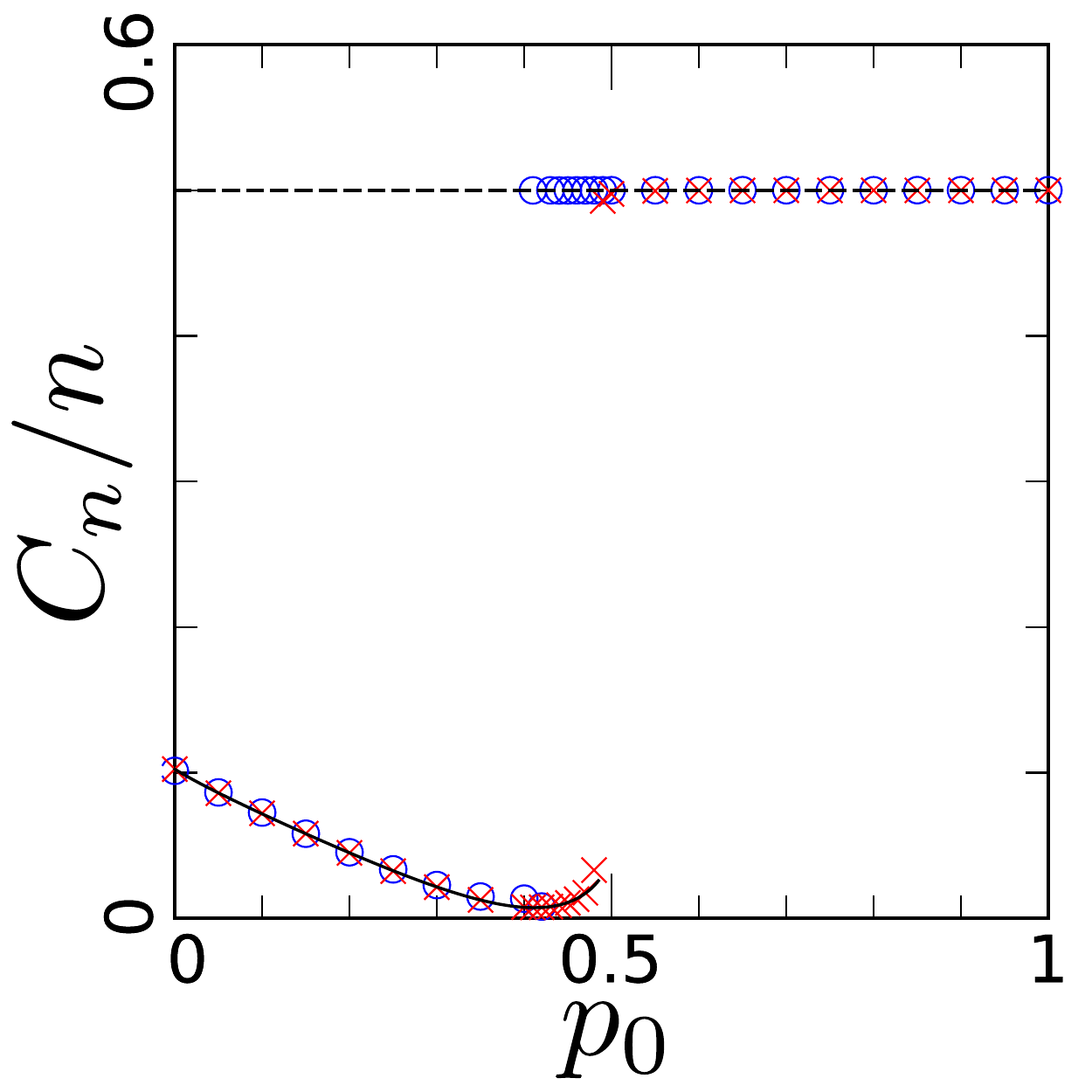} &
 \includegraphics[width=40mm]{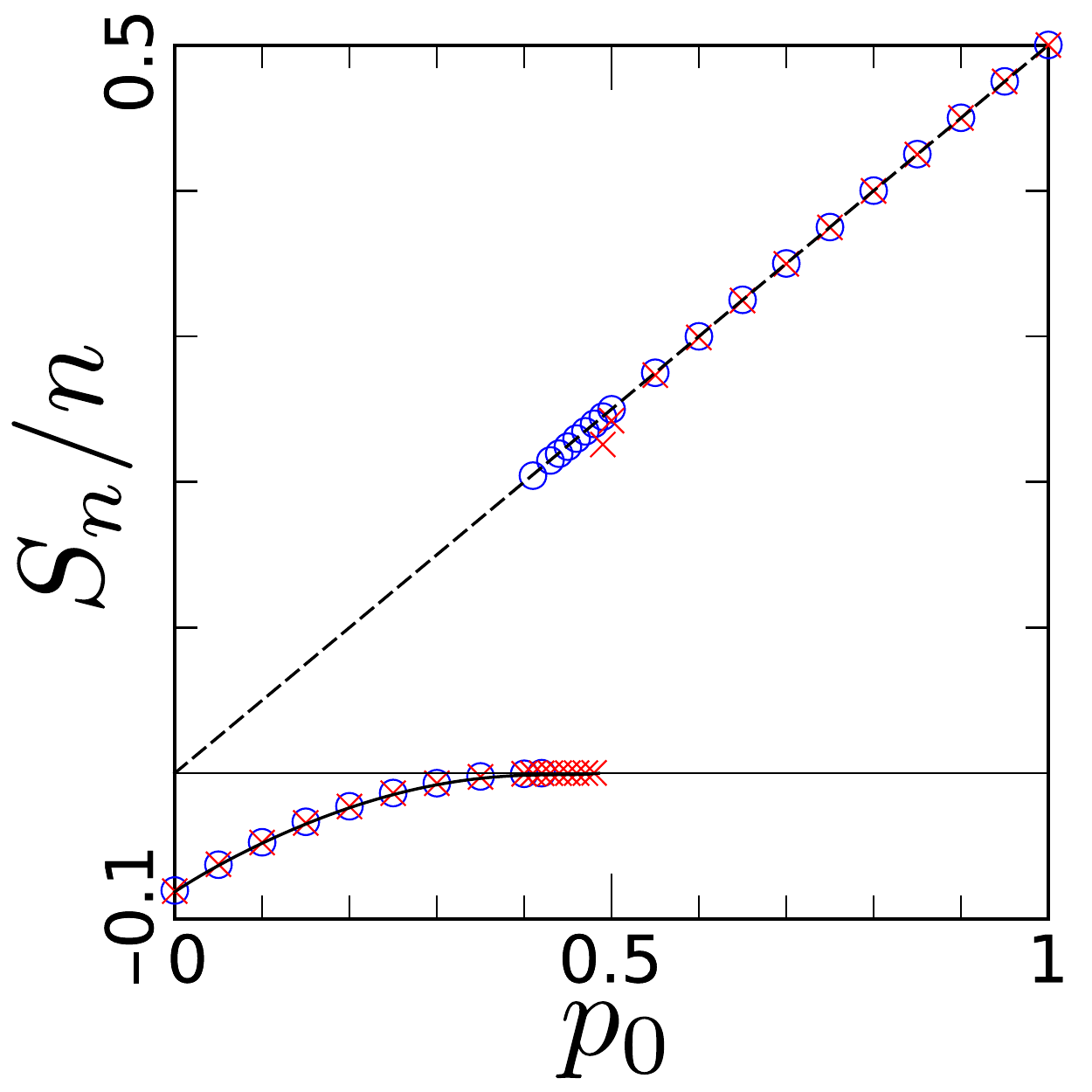} & 
 \includegraphics[width=40mm]{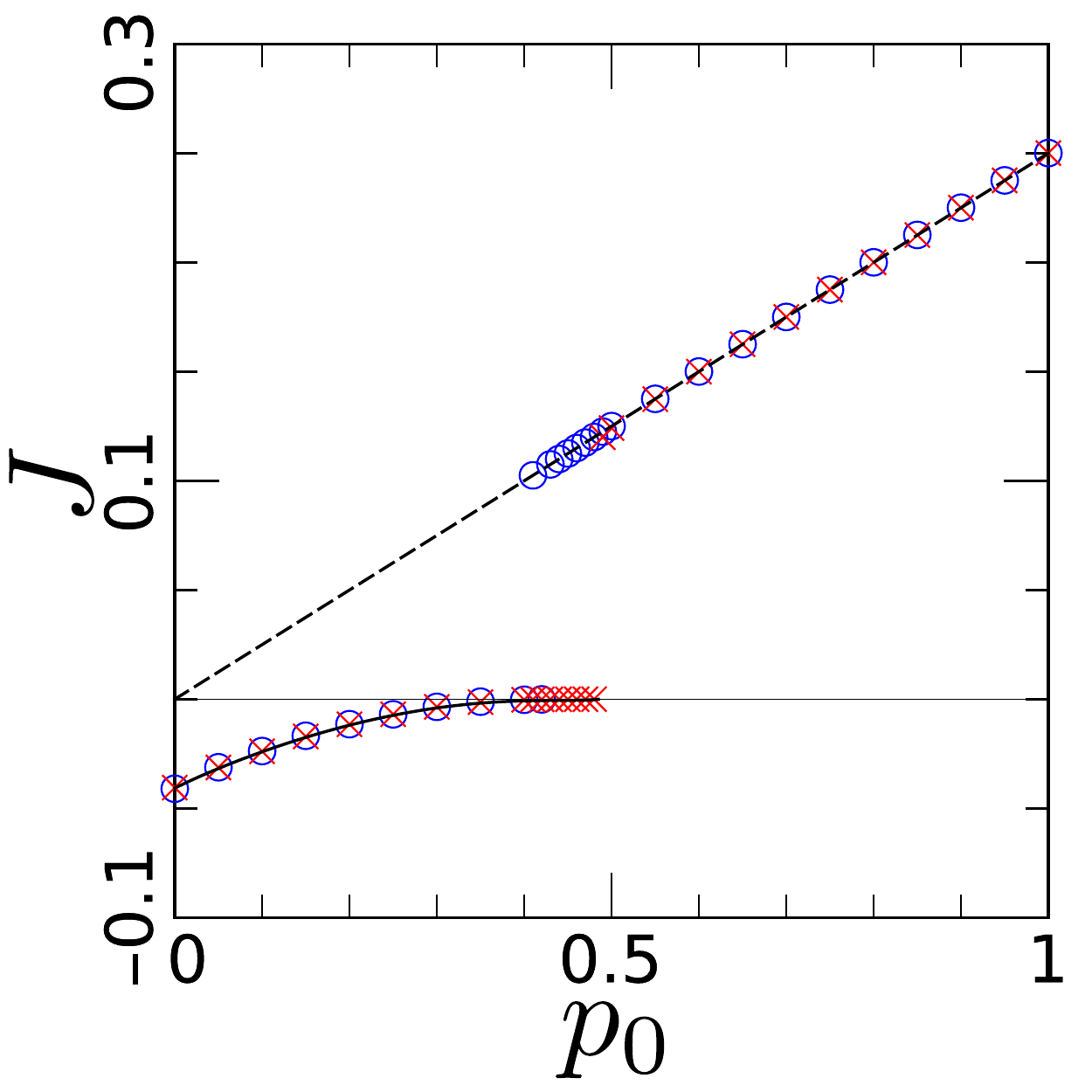} &
 \includegraphics[width=40mm]{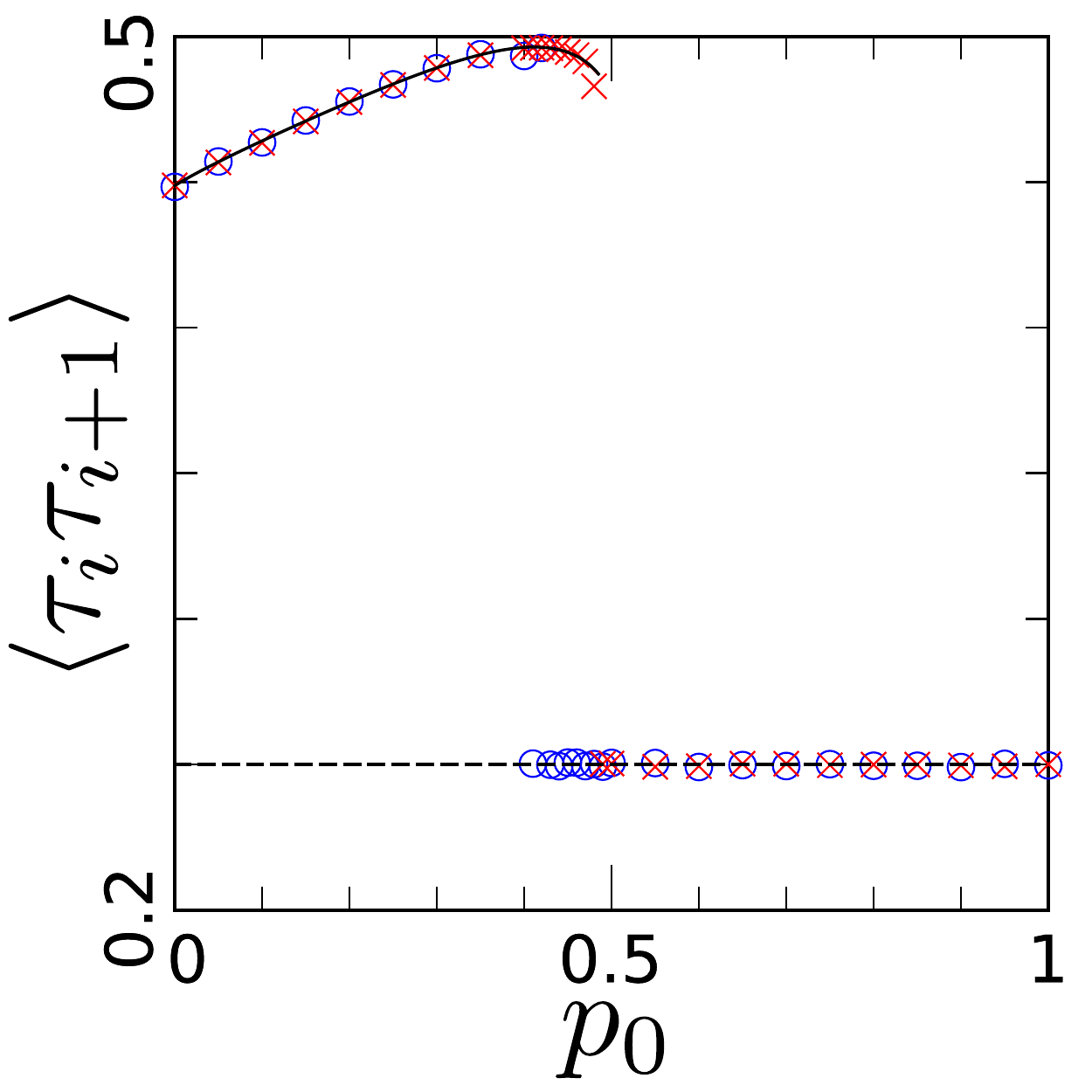}
\end{matrix} 
\end{align*}
 \caption{\label{fig:11}
 Simulation results of the quantity $ \frac{C_n}{n} $ (a), $ \frac{S_n}{n} $ (b), the current (c), and the correlation $ \langle \tau_i \tau_{i+1} \rangle $ (d) for $ p_\pm =1 $ and $ \rho=0.5 $. We used a chain with $L=4000$ sites, and imposed the step and random initial conditions, corresponding to different markers $ \times $ and $ \circ $. The dashed lines are the predictions \eqref{eq:C-mf}, \eqref{eq:S-mf}, \eqref{eq:J-mf} and \eqref{eq:tt-mf} of the global mean-field theory, and the solid lines correspond to the local mean-field formulas \eqref{eq:C-lf}, \eqref{eq:11-S-lf}, \eqref{eq:11-J-lf} and \eqref{eq:tt-lf}. 
}
 \end{center}
 \end{figure}

 When $p_0$ is large, we observe the distribution of particles is spatially uniform in the kymograph, see Fig.~\ref{fig:kymo} (a). Let us check the validity of the global mean-field formulas 
 \eqref{eq:tt-mf}, \eqref{eq:C-mf}, \eqref{eq:S-mf} and \eqref{eq:J-mf}.
 Note that, when $ p_\pm = 1 $, \eqref{eq:S-mf} and \eqref{eq:J-mf} are simplified as 
\begin{align}
\label{eq:11-S-mf}
 \frac{ S_n ^{ (k) } }{ n}
 & = \big[ 1 - 2 ( 1 - p_0 ) \rho \big] (1-\rho ) , \\
\label{eq:11-J-mf}
 J & = \big[ 1 - 2 ( 1 - p_0 ) \rho \big] \rho (1-\rho ) . 
\end{align}
In Fig.~\ref{fig:11}, we compare them with simulation results. We find that, for large values of $ p_0 $, 
the global mean-field formulas actually agree with simulations. 

On the other regime ($ p_0 $ is small), the global mean-field theory fails. In contrast to the uniform phase, there is a cluster of particles, see the kymograph, Fig.~\ref{fig:kymo} (b). Qualitatively, one can interpret the figures as follows. Since $p_+=p_-=1$, it implies that jumps forward are highly favored, except when $q_0=1-p_0$ is big. Then, for small values of $p_0$, for any group of particles, most of the particles will first stop (because of the exclusion principle) and then jump backward (since $q_0$ is high). As a consequence, the group will be slowed or even move backward, favoring the emergence of a cluster. Isolated particles will always jump forward (or stay).

 It explains the small negative values of $J$ for small values of $p_0$. Since in the cluster the particles do not move most of the time, it explains also the small value of $C_n/n$. Now, if $p_0$ becomes bigger than $q_0$ (i.e. $0.5\lesssim p_0$) the jumps forward are always favored, hence $J>0$, and linear in $ p_0 $, as we see in Eq.~\eqref{eq:11-J-mf}. A group of particles moves roughly in the same way as isolated particles, so that there is no clustering.

So far we have no rigorous argument to precisely predict the current in this condensed phase,
but we can use the local mean-field formulas \eqref{eq:C-lf}--\eqref{eq:tt-lf}. 
When $ p_\pm=1 $, we have simplifications for $ \frac{ S^{ (k) }_n}{n} $ and $ J $ as 
\begin{align}
\label{eq:11-S-lf}
 \frac{ S^{ (k) }_n}{n} & = [ 1 - (1- 2 p_0 ) \rho' ] (1-\rho' ), \\ 
 J & = [ 1 - (1- 2 p_0 ) \rho' ] (1-\rho' ) \rho. 
\label{eq:11-J-lf}
\end{align}
Recall that these formulas contain the cluster density $ \rho' $ as a free parameter. 
We start with looking for a good function $ \rho' ( p_0 ,\rho) $ 
 which reproduce the curve of $C_n/n = 1-\rho'$,
as shown in Fig.~\ref{fig:11}~(a). For example 
\begin{align}\label{eq:fitting-11}
& \rho' 
= \frac{a+b p_0-\sqrt{\alpha+\beta p_0+\gamma p_0^2+\delta p_0^3}}{1-p_0}, \\
& a\approx 0.88\,, 
\quad
 b\approx - 0.83\,,
 \quad
 \alpha\approx 0.00037\,,
 \quad
 \beta\approx 0.012\,, 
 \quad 
 \gamma\approx 0.016\,, 
 \quad
 \delta\approx 0.080\,. 
\end{align}
We remark that this form does not depend on the density $ \rho $. 
Then, we substitute the form \eqref{eq:fitting-11} into
\eqref{eq:11-S-lf}, \eqref{eq:11-J-lf} and \eqref{eq:tt-lf}. 
Now we have obtained predictions for $ S_n /n $, $ J$ and $\langle \tau_i\tau_{i+1} \rangle $,
which fit the simulation results very well, see Fig.~\ref{fig:11} (b)--(d).

\subsection{Phase transition}

One notices that the global mean-field predictions and the local mean-field predictions cannot have the same values in general. For example, for $ \rho=0.5 $, apparently the dashed and solid lines have no intersection, see Fig.\ref{fig:11}. Supported by this observation, we conjecture that our model exhibits first order phase transitions. 

Another interesting observation is initial-condition dependence. 
For the simulations, we used the following two different initial conditions: 
i) a step initial condition $11\cdots100\cdots0$, i.e. we put particles at sites $ 1 \le i \le N $
and ii) a randomly chosen configuration, 
which is equivalent to the Bernoulli measure with density $ \rho=N/L $ as $ L\to \infty $. 
 In Fig.~\ref{fig:11}, different plot markers are used in order to distinguish between these two initial conditions. 
We notice that that the phase transition point depends on the initial conditions, as it can be seen when looking closely at the plots. In other words, the ergodicity is broken in our model: for $ 0.4 \lesssim p_0 \lesssim 0.5 $, the system can take both uniform and condensed configurations in the long time limit, depending on the initial conditions.

\section{Case $ p_+= 0 , \ p_-= 1 $}\label{sec:01}

\subsection{Uniform and condensed phases}

In the case of $ ( p_+ , p_- ) = ( 0,1 ) $, Eq. \eqref{eq:P} becomes 
\begin{align}
 P = p_{0} - \frac{ 1 }{2(n-1)} S_{n-1}^{(k)} + \frac{ 1 - 2p_{0} }{2(n-1)} C_{n-1}^{(k)} . 
\end{align}
We remark that the symmetry \eqref{eq:flip} reads $ J ( p_0,\rho ) = - J ( 1-p_0,\rho ) $ in this case. 
The global mean-field formulas are simplified as 
\begin{align}
\label{eq:01-snn-mf}
 \frac{1}{n}S_n &= (2 p_0 - 1) \frac{ \rho(1-\rho) }{ 2-\rho } , \\
 J &= (2p_0-1) \frac{ \rho^2 (1-\rho) }{ 2-\rho } . 
\label{eq:01-J-mf}
\end{align}

\begin{figure}[h]
 \begin{center} 
\begin{align*}
\begin{matrix} 
 \textbf{(a)} & \textbf{(b)} & \textbf{(c)} & \textbf{(d)} \\ 
 \includegraphics[width=40mm]{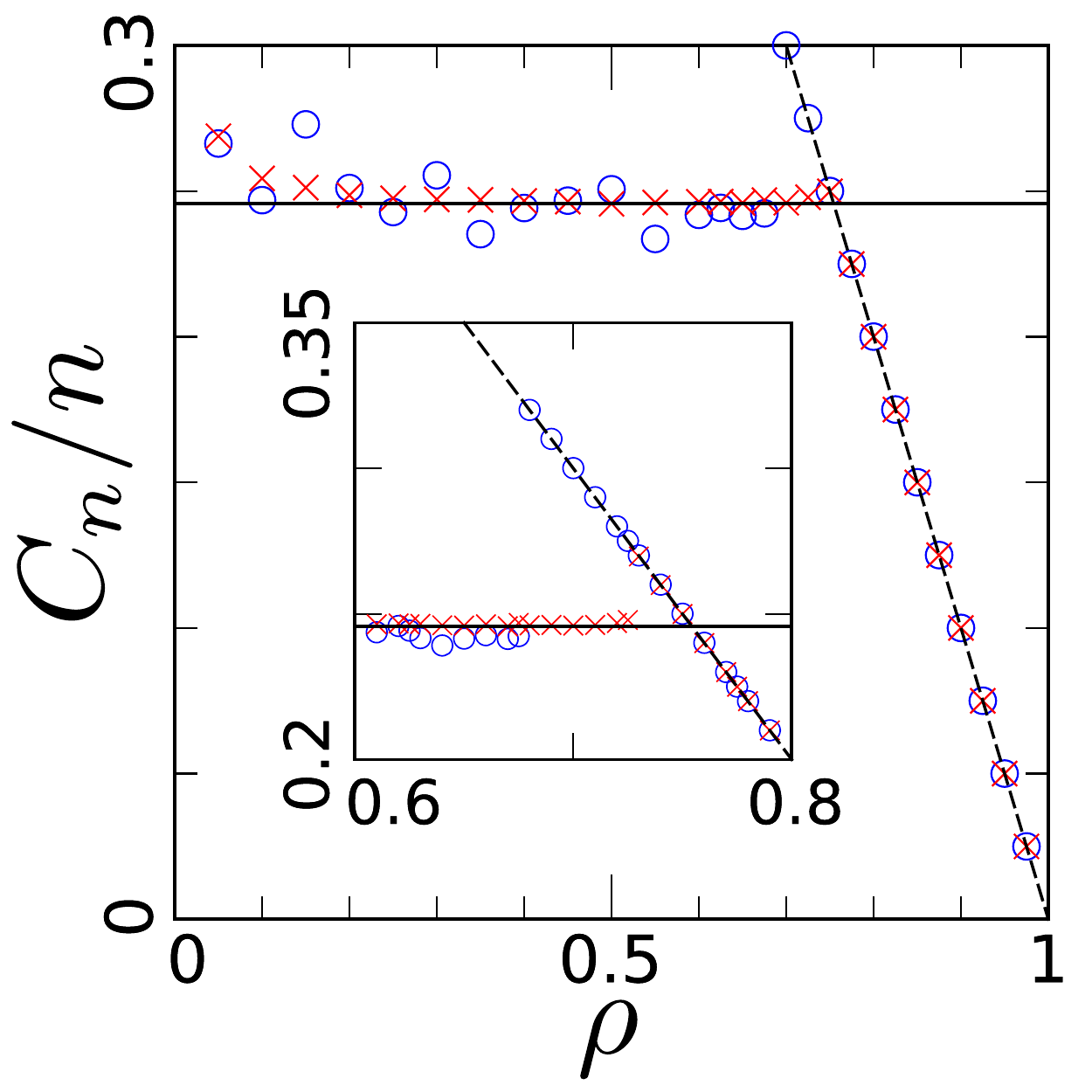} & 
 \includegraphics[width=40mm]{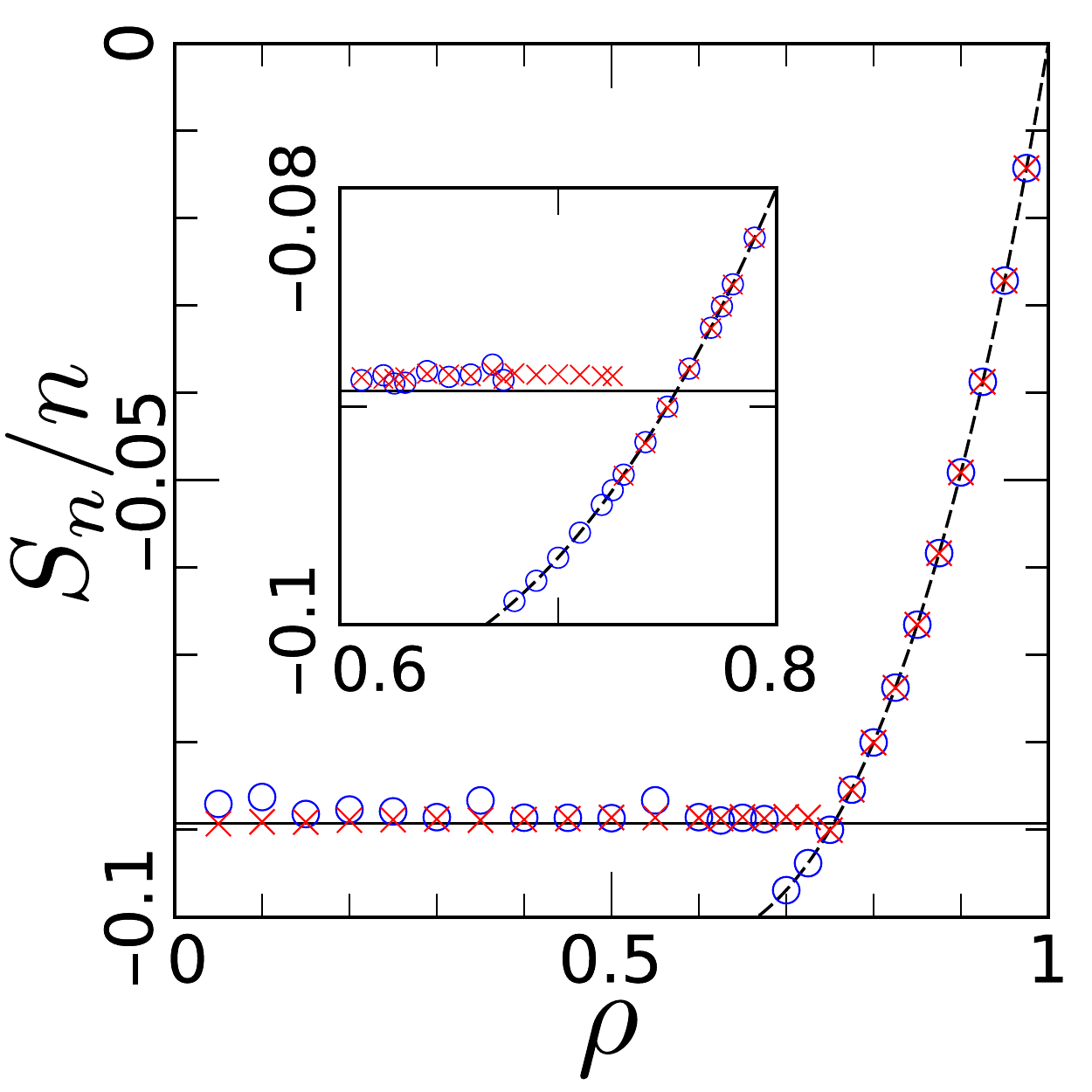} & 
 \includegraphics[width=40mm]{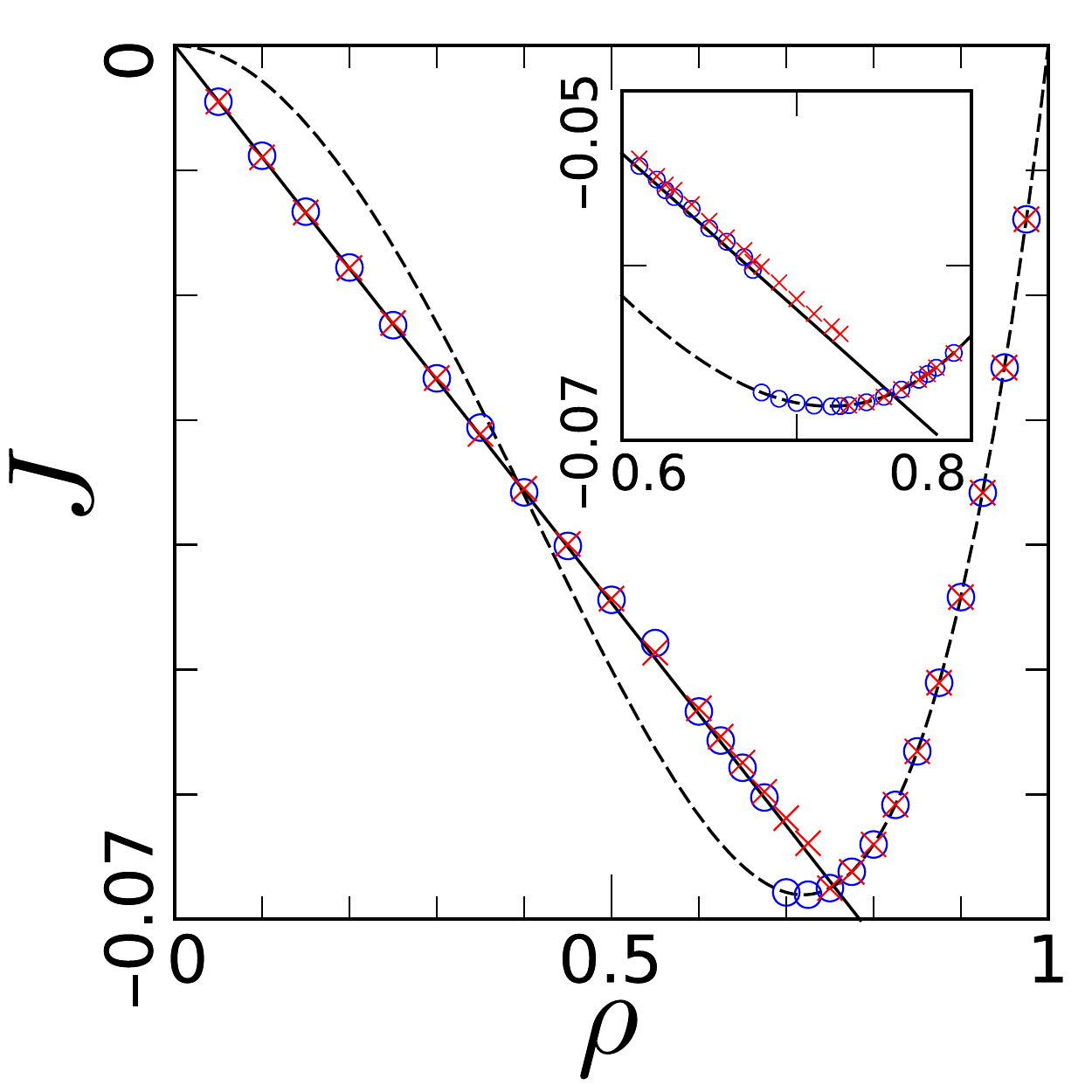} & 
 \includegraphics[width=40mm]{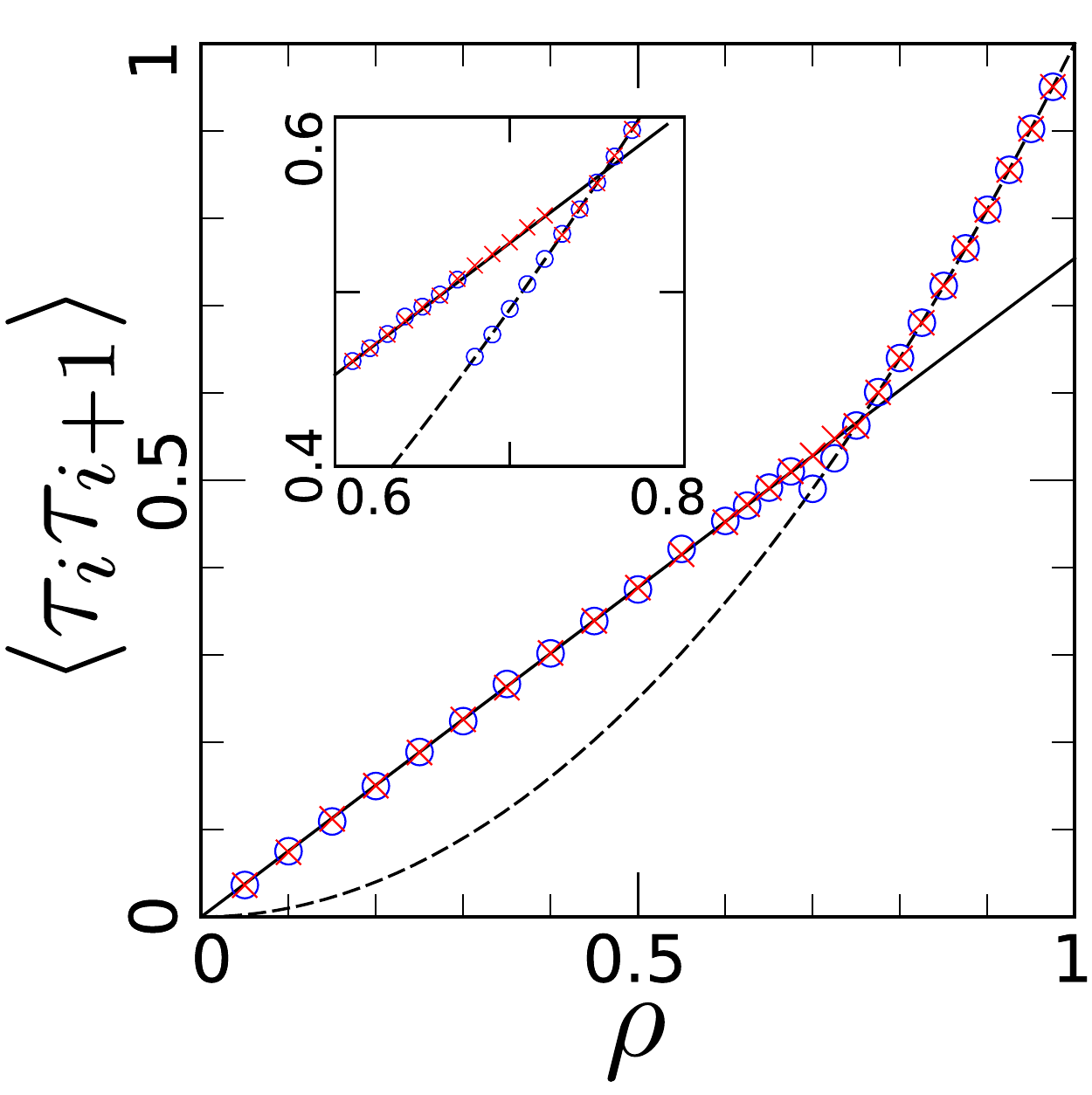}
\end{matrix} 
\end{align*}
 \caption{\label{fig:01}
 Simulation results of (a) the quantity $ \frac{C_n}{n} $, (b) the quantity $ \frac{S_n}{n} $, (c) the current, and (d) the correlation $ \langle \tau_i \tau_{i+1} \rangle $ for $ p_+ = 0$, $p_- = 1 $ and $ p_0 = 0.2 $. We used a chain with $L=4000$ sites, and imposed the step and random initial conditions, corresponding to different markers $ \times $ and $ \circ $. 
 The global mean-field predictions (dashed lines) for the four panels are $ \frac{C_n}{n} = 1-\rho $,
 Eq.~\eqref{eq:01-snn-mf}, Eq.~\eqref{eq:01-J-mf}, and $ \langle \tau_i \tau_{i+1} \rangle = \rho^2 $. 
 In addition, the solid lines correspond to the local mean-field predictions 
 \eqref{eq:c-01-lf}--\eqref{eq:tt-01-lf}.
 In each panel, an inset is given in order to closely look at the data
 near the phase transition. 
}
 \end{center}
 \end{figure}

In Fig.~\ref{fig:01}, we show simulation results for $ C_n /n $, $ S_n /n $, 
the current and the correlation, as functions of the density $ \rho $. 
As we found in the previous case, there are two different regimes, 
but for the present case the phase depends on $\rho$ rather than $p_0$.

We observe good agreements between the global mean-field formulas and simulations 
when $ \rho $ is large (say, $\rho> \rho^* $), see Fig.~\ref{fig:01}.
In the kymograph, Fig.~\ref{fig:kymo}~(c), we find that the particles are uniformly distributed. 
On the other hand, for $ \rho < \rho^* $, the global mean-field prediction fails, and a cluster is spontaneously formed, see the kymograph, Fig.~\ref{fig:kymo}~(d). Apparently the particle distribution is not uniform over the chain. 
The critical value $\rho^*$ can be qualitatively interpreted in the following way. When $\rho$ is high, most of the particles are gathered and cannot move. Then, the jumps are governed by $p_0$ and $\rho$ (since the particles need some `space' to jump), as given by the global mean field approximation. On the contrary, for small concentrations, the values $p_-=1$ and $p_+=0$ imply that in average each jump backward is followed by a jump forward (and vice-versa), so that in average $S_n/n$ is small (its non-zero value ruled by $p_0$ and independent from $\rho$).

Now, we use the local mean-field theory in the condensed phase. In Fig.~\ref{fig:01}~(a) we observe that $ \frac{C_n}{n} $ is constant. Equivalently we conjecture that $ \rho' $ is constant,
 and numerically we have 
\begin{align}
 \frac{C_n}{n} = 1- \rho' \approx 0.25 , \quad 
 \rho' \approx 0.75 . 
 \label{eq:c-01-lf}
\end{align}
Our simulations for different values of $ p_0 $ (see Appendix) indicate that the cluster density $ \rho' $ is also constant for $ p_0 $. 
Then the other quantities should be 
\begin{align}
 \frac{S_n}{n} & = (2 p_0 - 1) \frac{ \rho' (1-\rho') }{ 2-\rho' } \approx 0.15 (2p_0-1) , \\ 
 J & = \rho (2p_0-1) \frac{ \rho' (1-\rho') }{ 2-\rho' }
 \approx 0.15 (2p_0-1) \rho , \\ 
\langle \tau_i \tau_{i+1} \rangle & = \rho \rho' 
\approx 0.75 \rho . \label{eq:tt-01-lf}
\end{align}
These expressions fit very well the simulation results in the condensed phase, see Fig. \ref{fig:01}.

\subsection{Phase transition}

In the case of $ p_+= 0 , \ p_-= 1 $, the cluster density $ \rho' $ is observed to be constant \eqref{eq:c-01-lf}. 
Apparently the global density $ \rho $ cannot exceed $ \rho' $, and one may expect that $ \rho' $ gives the transition point. In other words one could guess that a second order phase transition occurs just at the intersection of the dashed line and the solid line in each panel of Fig.~\ref{fig:01}. However, this naive consideration is wrong, as we closely see the insets. The actual transition point is lower than $ \rho' $ and it depends on the initial conditions. Furthermore, we find that the phase transitions are of first order.

\section{Discussions}\label{sec:discussions}

In this work, we introduced a system of interacting elephant random walks on a ring. We imposed the exclusion principle, i.e. each site can be occupied by at most one particle. Therefore our model contains the exclusion process as a special case. We performed Monte Carlo simulations, and found that interacting elephants exhibit two different behaviors. One is the uniform phase, where particles are uniformly distributed over the chain, and the other is the condensed phase, where a cluster is formed. For the former case, the current is easily predicted by neglecting correlations, to which we refer as the global mean-field theory. For the latter case, we introduced a local mean-field theory, assuming that the particles are uniformly distributed inside the cluster.

The condensation phenomenon of our system is induced by a different mechanism from another non-Markovian exclusion process \cite{bib:CB}. The authors of \cite{bib:CB} imposed a time series of particle jumps obeying a power-law tail $ \tau^{-\gamma} $ for each particle, which induces a condensation. When the waiting time between an attempt of jump of a particle and the next attempt is very long, a long queue is formed before this particle, i.e. a \textit{static} cluster appears. Therefore condensed and uniform states in their model are distinguished by tuning the exponent $\gamma$ (see also \cite{bib:KHG}). Contrarily, in our system the time series of particle jumps is given by exponential distribution as in the case of the standard exclusion process, but the direction depends on the history. The behavior of our model is determined by the directionality parameters $ p_\pm ,p_0$ and the density $ \rho $.
 We also mention a zero-range process with memory effect in one dimension \cite{bib:HMS}, where each site is allowed to be occupied by more than one particle. This model exhibits another type of condensation, i.e. two sites are occupied by particles but the rest of the sites are empty, and the group of particles moves like a ``slinky''. 

 Our simulations indicate that the phase transition is of first order. It is also remarkable that the initial configuration affects the phase in the long time limit. 
Our local mean-field theory contains an extra parameter $ \rho' $ which is the particle concentration in the cluster. So far we have obtained the value of $ \rho' $ only numerically and by performing fitting. We also found that the phase transition point is not given by $ \rho=\rho' $. Clearly, a more theoretical study of the condensed phase and of the phase transition is deserved to be done.

\section*{Acknowledgements}
This work was supported by the grant AAP MASHE from the Savoie Mont Blanc University.

\section*{Appendix}

 \begin{figure} [h]
 \begin{center} 
\begin{align*}
\begin{matrix} 
 \textbf{(a)} & \textbf{(b)} & \textbf{(c)} & \textbf{(d)} \\ 
 \includegraphics[width=40mm]{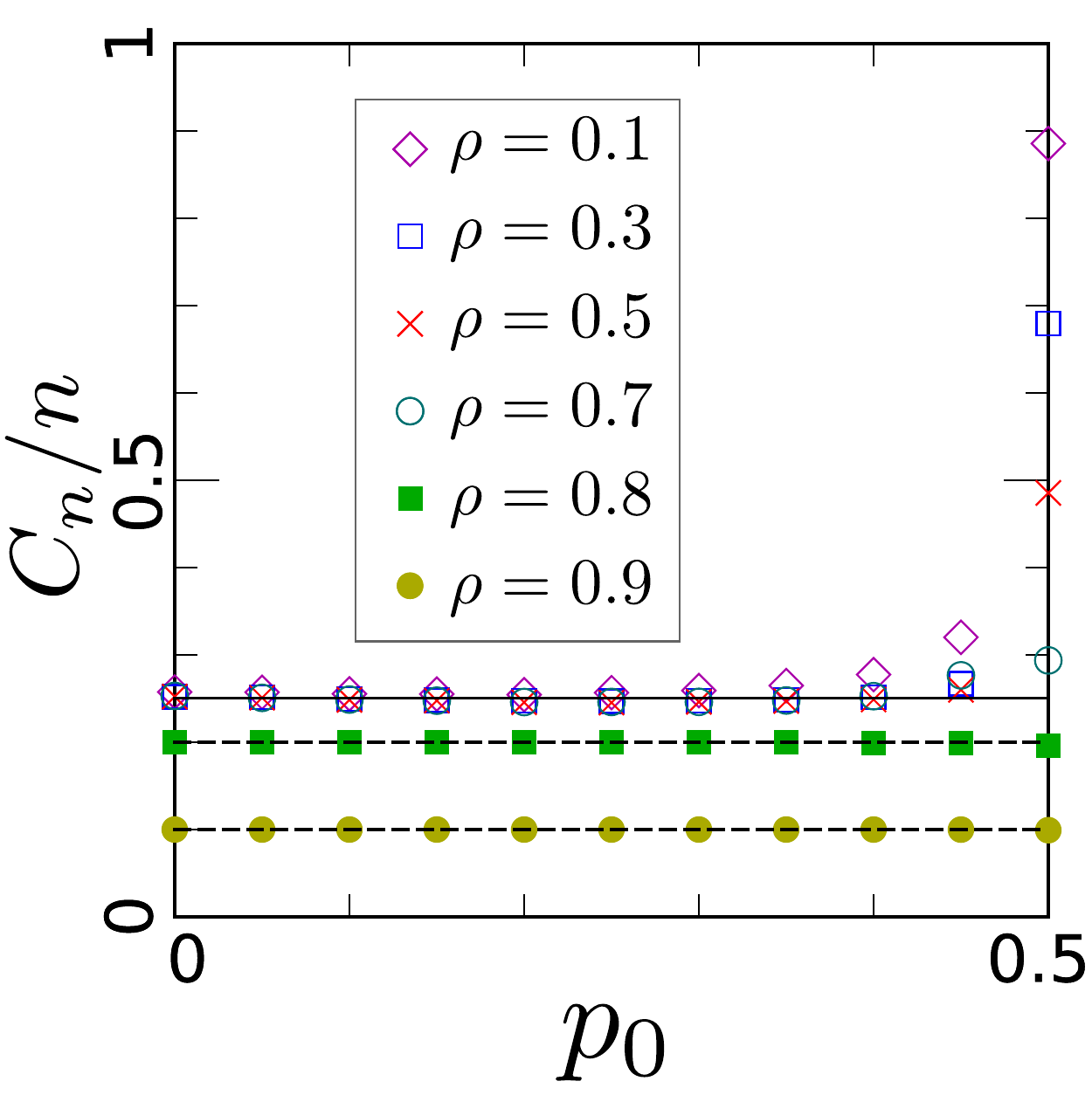} &
 \includegraphics[width=40mm]{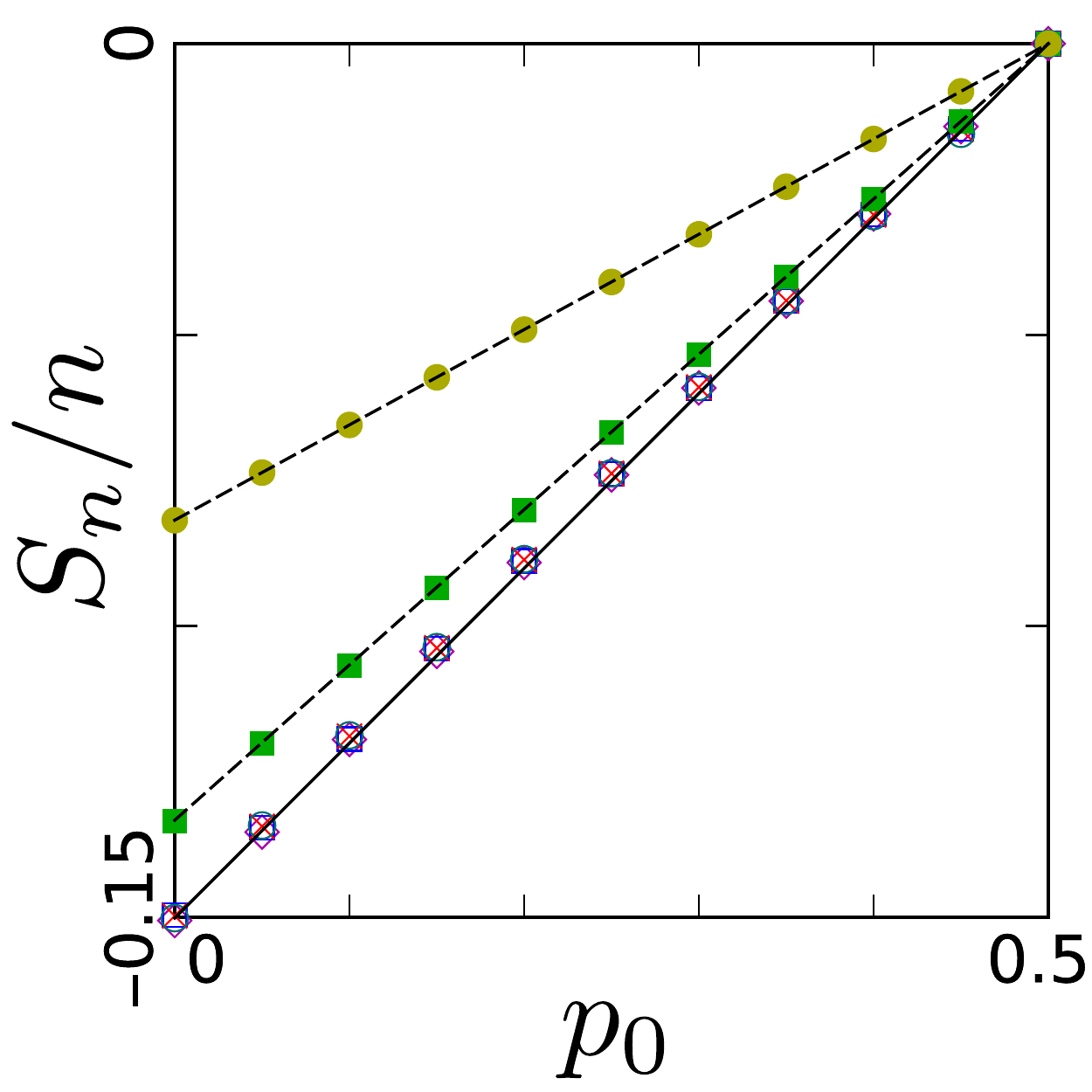} &
 \includegraphics[width=40mm]{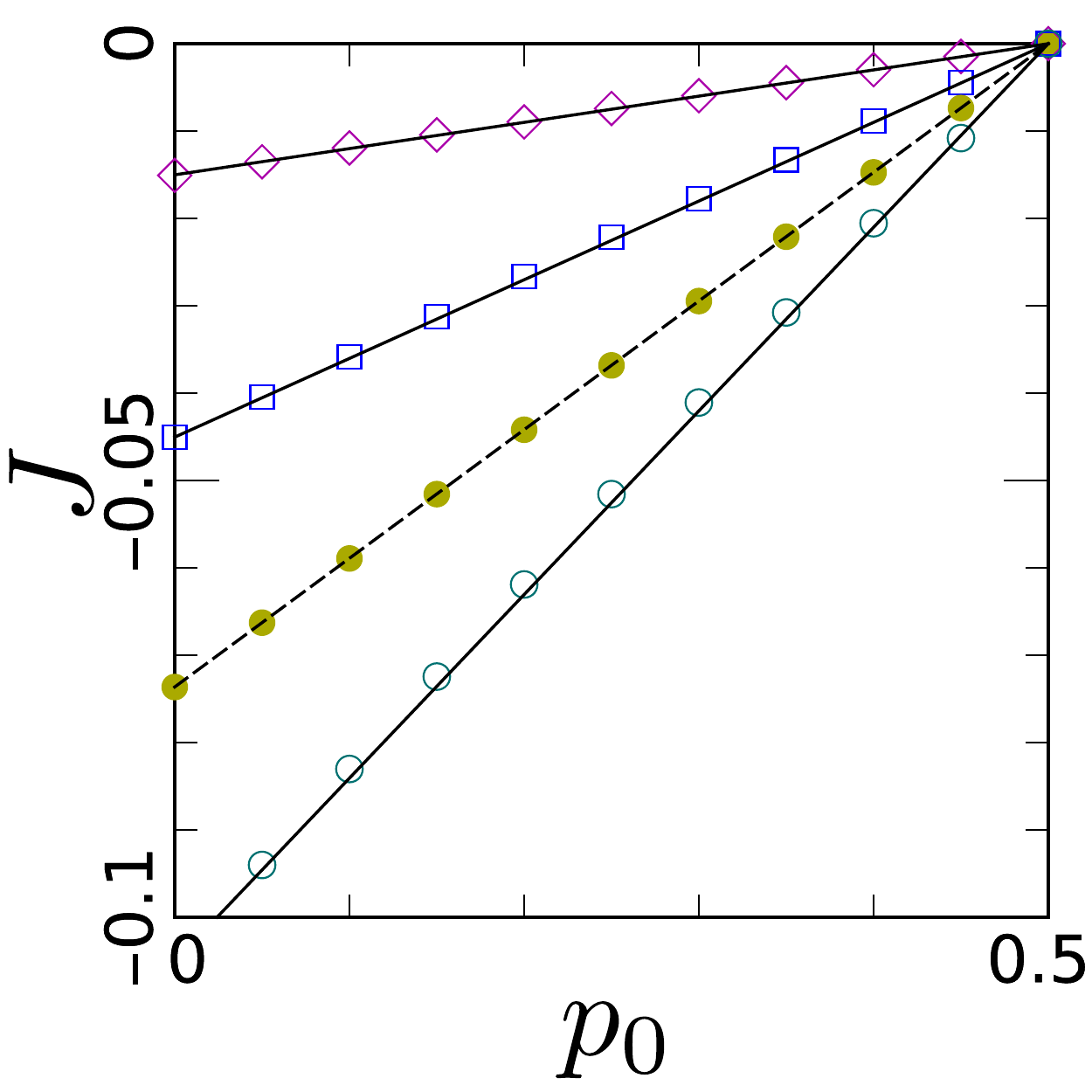} &
 \includegraphics[width=40mm]{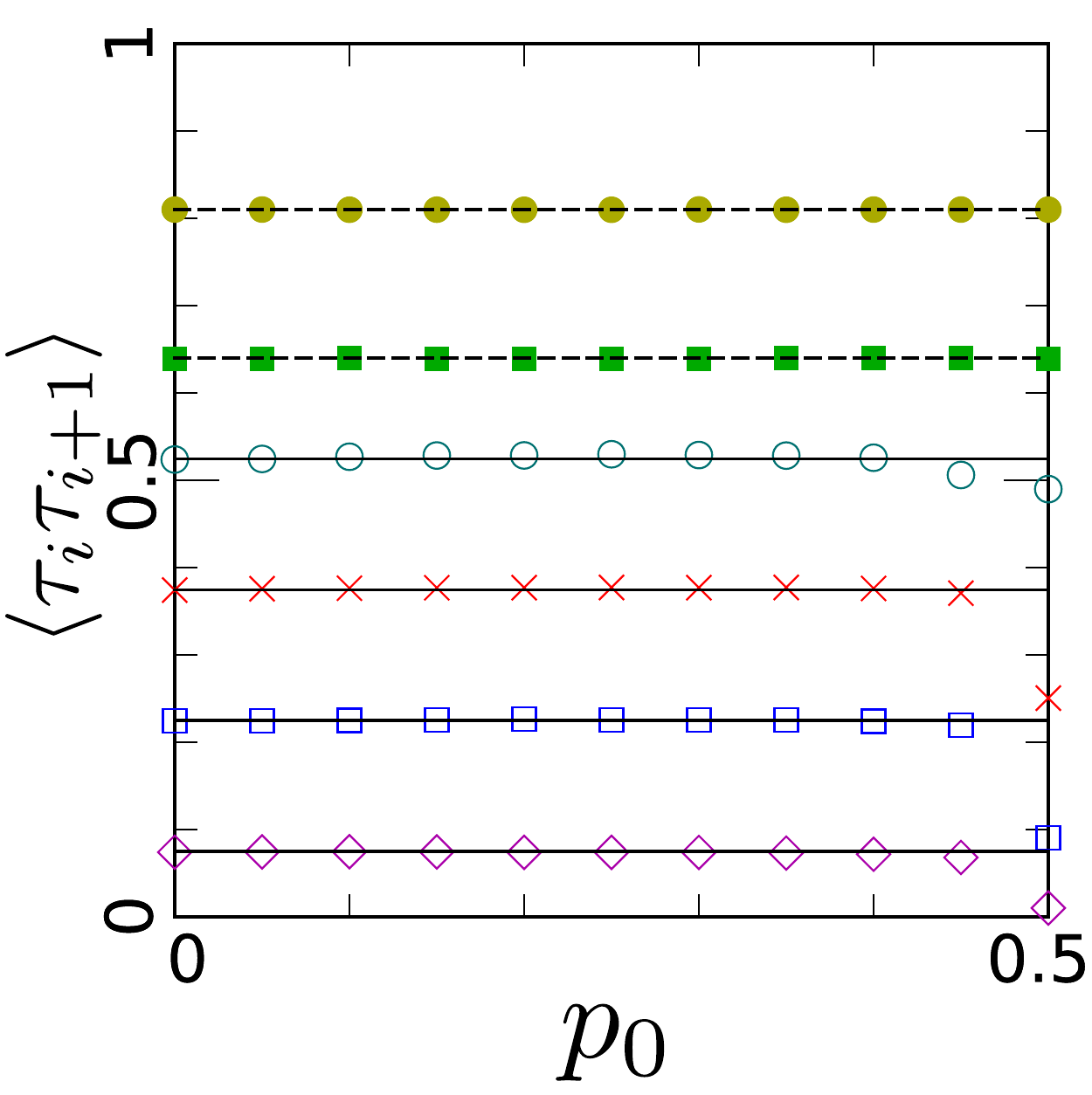}
\end{matrix} 
\end{align*}
 \caption{\label{fig:01-app}
 Simulation results of (a) the quantity $ \frac{C_n}{n} $, (b) the quantity $ \frac{S_n}{n} $, (c) the current, and (d) the correlation $ \langle \tau_i \tau_{i+1} \rangle $ vs $ p_0 $
 for $ p_+ = 0$, $p_- = 1 $. We used a chain with $L=4000$ sites, and imposed the step initial condition. According to the global density $ \rho $, different plot markers are used. 
 The global mean-field predictions (dashed lines) in the four panels 
 are $ \frac{C_n}{n} = 1-\rho $,
 Eq.~\eqref{eq:01-snn-mf}, Eq.~\eqref{eq:01-J-mf}, and $ \langle \tau_i \tau_{i+1} \rangle = \rho^2 $
 for $ \rho > \rho^* \approx 0.73 $. 
 The local mean-field predictions \eqref{eq:c-01-lf}--\eqref{eq:tt-01-lf} for $ \rho < \rho^* $ 
 are shown by the solid lines. 
}
 \end{center}
 \end{figure}

\noindent 
We plot simulation results for $ p_+ = 0$, $p_- = 1 $ as functions of $ p_0 $ in Fig.~\ref{fig:01-app}. Since we have the symmetries 
\begin{align}
 C_n/n \big|_{ p_0 \to 1-p_0 } = C_n/n, \quad 
 S_n/n \big|_{ p_0 \to 1-p_0 } = -S_n/n, \quad 
 J \big|_{ p_0 \to 1-p_0 } = -J, \quad 
 \langle \tau_i\tau_{i+1}\rangle \big|_{ p_0 \to 1-p_0 } = \langle \tau_i\tau_{i+1}\rangle , 
\end{align}
we only show the data in the range $ 0 \le p_0\le 0.5 $. Figure~\ref{fig:01-app}~(a) indicates that the cluster density $ \rho' $ is always $ 1-C_n/n\approx 0.75 $ in the condensed phase ($ \rho < \rho^* \approx 0.73 $), except when $ p_0 $ is near $0.5$. The correlation function (d) is also almost constant, but the quantities $ S_n/n $ (b) and $ J $ (c) are linear in $ p_0 $. The point $ p_0=0.5 $ is special: independently from the global density $ \rho $, the model converges to the uniform phase.


\clearpage

\begin{thebibliography}{99}

\bibitem{bib:ST} G.~M.~Sch\"utz and S.~Trimper, Phys. Rev. E 70, 045101 (R) (2004).

\bibitem{bib:PE} F.~N.~C.~Paraan and J.~P.~Esguerra, Phys. Rev. E 74, 032101 (2006). 

\bibitem{bib:MGP} C.~T.~ MacDonald, J.~H.~Gibbs, and A.~C.~Pipkin, Biopolymers 6, 1 (1968).

\bibitem{bib:Spitzer} F.~Spitzer, Adv. Math. 5, 246 (1970). 

\bibitem{bib:KRB-N} 
P.~L.~Krapivsky , S.~Redner, and E.~Ben-Naim, A Kinetic View of Statistical Physics (Cambridge: Cambridge
University Press, 2010).

\bibitem{bib:CMZ} 
T.~Chou, K.~Mallick, and R.~K.~P.~Zia, Rep. Prog. Phys. 74, 116601 (2011).

\bibitem{bib:Schuetz} G.~M.~Sch\"utz, 
Exactly solvable models for many-body systems far from equilibrium,
Phase Transitions and Critical Phenomena vol 19, ed C Domb and J L Lebowitz
 (San Diego, Academic, 2001).

\bibitem{bib:CSV} J.~C.~Cressoni, M.~A.~A.~da Silva, G.~M.~Viswanathan, Phys. Rev. Lett. 98, 070603 (2007).
 
\bibitem{bib:KHL} N.~Kumar, U.~Harbola, and K.~Lindenberg, Phys. Rev. E 82, 021101 (2010).

\bibitem{bib:CB} 
R.~J.~Concannon and R.~A.~Blythe, Phys. Rev. Lett. 112 050603 (2014).

\bibitem{bib:KHG} 
 D.~Khoromskaia, R.~J.~Harris, and S.~Grosskinsky, J. Stat. Mech. (2014) P12013.

\bibitem{bib:HMS} 
O.~Hirschberg, D~Mukamel and G.~M.~Sch\"utz, Phys. Rev. Lett. 103, 090602 (2009); 
 J. Stat. Mech. (2012) P08014.

\end{thebibliography}
\end{document}